\documentclass[twocolumn,amsmath,amssymb,pra]{revtex4}
\usepackage{setspace}
\usepackage{graphicx}
\usepackage{amsmath}
\usepackage{amssymb}
\usepackage{latexsym}

\usepackage{natbib}
\begin{document}
\title{Spontaneous creation of non-zero angular momentum modes in tunnel-coupled two-dimensional degenerate Bose gases}
\author{T.W.A. Montgomery, R.G. Scott, I. Lesanovsky, and T.M. Fromhold}
\affiliation{Midlands Ultracold Atom Research Centre, School of Physics and Astronomy, University of Nottingham, Nottingham, NG7 2RD, United Kingdom.}
\date{07/10/09}


\begin{abstract} 
We investigate the dynamics of two tunnel-coupled two-dimensional degenerate Bose gases. The reduced dimensionality of the clouds enables us to excite specific angular momentum modes by tuning the coupling strength, thereby creating striking patterns in the atom density profile. The extreme sensitivity of the system to the coupling and initial phase difference results in a rich variety of subsequent dynamics, including vortex production, complex oscillations in relative atom number and chiral symmetry breaking due to counter-rotation of the two clouds.
\end{abstract}

\maketitle

\section{Introduction}

A major focus of cold-atom research is coupling multiple degenerate Bose gases (DBGs) using atom chips and optical lattices~\cite{krugerRev,reichelrev,fortaghrev,kasevich,cornell,kettnew,kettnew2,schmiedreview,Hofferberth,andreasdiff2,mott,JJExper,kruger,OberPRL}. The results provide stepping stones to future applications, such as interferometers or processors of quantum information. Since atom chips and optical lattices typically generate very high (kHz) trapping frequencies, there is growing interest in the dynamics of coupled one- and two-dimensional (1D and 2D) clouds~\cite{savin,annular,bouchoule,brandfluxons,malomed}. This interest has also been fueled by the radically different physics that has been observed in lower-dimensional systems, such as the suppression of equilibration~\cite{newtons}, quasicondensation~\cite{phasedefects}, and the Kosterlitz-Thouless transition~\cite{hello,kruger07PRL,kruger08NJP,CornellPRL,PhillipsPRL,TapioPRL,DalibardRev}. Since this body of work has uncovered such rich dynamics, it is natural to wonder how coupled lower-dimensional systems will behave, and whether they can reveal a crossover to 3D phenomena. Some recent work on 1D coupled rings has shown that the reduced dimensionality leads to unexpected effects, such as spontaneous population of rotating excitations and chiral symmetry breaking~\cite{annular}. Further work is now needed to explore symmetry breaking in Josephson junctions, the boundaries of lower-dimensional physics, and to establish how double-well interferometers will perform in reduced dimensions.



In this paper, we investigate the dynamics of coupled 2D disk-shaped DBGs. We find that the reduced dimensionality of the clouds has profound implications for the dynamics, because the instabilities of excited states observed in three dimensions are suppressed~\cite{RScottInter,RScottInter2}. Starting from an irrotational stationary state, and without introducing any stirring, we observe spontaneous occupation of low-lying excitations with non-zero angular momentum, mediated by the interatomic interactions~\cite{annular,spinchiral,saito}. We use a linear stability analysis to predict which rotating Bogoliubov modes will be excited, and hence identify regimes where we can excite specific modes \textit{alone} by tuning the coupling strength. This targeted selection of a single rotating mode creates striking oscillatory patterns in the atomic density profile. As we excite different rotating modes, we uncover a rich parameter space. Modes with a high angular momentum periodically grow and decay exponentially. The growth rates of these modes are extremely sensitive to changes in the coupling and interaction strengths, and in the relative phase of the two DBGs. The growth of modes with lower angular momentum becomes unstable due to collisions between rotating and non-rotating atoms, disrupting the internal structure of the clouds. This leads to a variety of subsequent dynamics such as vortex production, oscillations in relative atom number, and chiral symmetry breaking due to counter-rotation of the two clouds. 


\section{System and Methodology}
\vspace{-0.2 cm}

\begin{figure}[tbp]
\includegraphics[width=0.91\columnwidth]{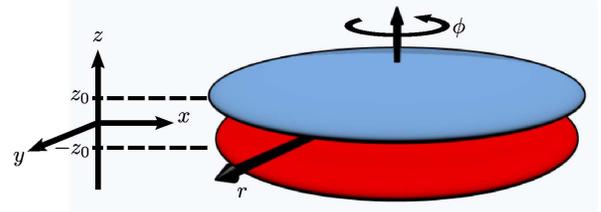} 
\vspace{-0.2 cm}
\caption{(Color online) Schematic constant density surfaces of the upper (blue/light grey) and lower (red/dark grey) DBGs. Arrows show co-ordinate axes.}
\label{f0}
\end{figure}

The system consists of two 2D DBGs, referred to as the upper and lower DBGs, in the $z = + z_{0}$ and $- z_{0}$ planes respectively, as shown in Fig~\ref{f0}. The two DBGs are coupled by a tunnel junction created by the symmetric double well potential $V_{\text{dw}}(z)$. In the $x-y$ plane, the DBGs are contained by the harmonic trapping potential $V(r) = \frac{1}{2} m\omega^{2} r^{2}$, where $m$ is the mass of a single atom, $\omega$ is the trap frequency and $r=\sqrt{x^2+y^2}$. For $\left|z\right|\approx z_{0}$, $V_{\text{dw}}(z)$ can be approximated by the tight harmonic potential $V_{\text{sw}}^{\alpha/\beta}(z) = \frac{1}{2} m \lambda^2 \omega^2(z \mp z_{0})^2$, where $\lambda \gg 1$ to ensure that $\lambda\hbar\omega > \mu$. Consequently, atomic motion in the $z$ direction is frozen into the single-particle groundstate, and the DBG wavefunction becomes 2D~\cite{2dness}. Hence, in the weak coupling limit \cite{JJTheo}, we may represent the order parameter for the two 2D DBGs by the scalar complex field
\begin{equation}\label{eq_ansatz1}
\psi = \zeta(z-z_{0})\chi^{\alpha}(\rho,\phi,\tau) + \zeta(z+z_{0})\chi^{\beta}(\rho,\phi,\tau)
\end{equation}
where the superscripts $\alpha$ and $\beta$ refer to the upper and lower DBGs, $\zeta$ is the normalized single-particle harmonic groundstate of $V_{\text{sw}}^{\alpha/\beta}(z)$, $\phi$ is the azimuthal angle, and we have introduced the dimensionless time $\tau = \omega t$ and the dimensionless length $\rho = r/a_{\text{ho}}$, in which $a_{\text{ho}} = \sqrt{\hbar/m\omega}$. Substituting Eq. \eqref{eq_ansatz1} into the Gross-Pitaevskii equation results in two coupled equations for $\chi^{\alpha/\beta}(\rho,\phi,\tau)$~\cite{foot2}
\begin{equation}\label{eq_gp2d}
\begin{array}{c c}
  i\partial_{\tau}\chi^{\alpha/\beta} & = -\left[\partial^2_{\rho} + \frac{1}{\rho}\partial_{\rho} + \frac{1}{\rho^2}\partial^2_{\phi} - \rho^2 +\mu\right]\chi^{\alpha/\beta}\\ & \ \ \  + \ \gamma|\chi^{\alpha/\beta}|^2\chi^{\alpha/\beta} - |\kappa|\chi^{\beta/\alpha}
\end{array}
\end{equation}
where the dimensionless quantities $\gamma = \left(8\pi\lambda\right)^{1/2}a_{0}/a_{\text{ho}}$ and $\kappa = \frac{1}{2} a^2_{\text{ho}}\int{\zeta(z+z_{0}) \left[\partial^2_{z}-2m V_{\text{dw}}(z)/\hbar\right] \zeta(z-z_{0})dz}$ represent the interaction and coupling energy respectively, and $a_{0}$ is the s-wave scattering length.

At $\tau=0$, there are an equal number of atoms, $N_{0}$, in each well, and $\chi^{\alpha/\beta}$ is the non-rotating groundstate of $V(r)$, with chemical potential $\mu_{0}$. For this initial configuration, and finite $\kappa$, there are two possible stationary states of Eq. \eqref{eq_gp2d}. These are the ground state, defined by $\chi^{\alpha}(r,\phi,0) = \chi^{\beta}(r,\phi,0)$, with chemical potential $\mu_{0} - |\kappa|$, and the excited asymmetric stationary state, henceforth referred to as the $\pi$-state, defined by $\chi^{\alpha}(r,\phi,0) = -\chi^{\beta}(r,\phi,0)$, with chemical potential $\mu_{0} + |\kappa|$~\cite{brandcomment}. Unsurprisingly, the ground state is stable for all coupling strengths. However, the $\pi$-state shows much richer behavior. In three dimensions, the phase discontinuity may bend, creating vortices via the well known snake instability~\cite{RScottInter,RScottInter2,dutton,anderson}. This process cannot occur in our system, because the reduced dimensionality of the disks precludes the movement of the phase discontinuity. In the following section, we perform a stability analysis to identify what excitations may occur. 

\section{Stability analysis of the $\pi$-state}

We perform a stability analysis of the stationary states by calculating the excitations of the system using the Bogoliubov ansatz~\cite{stability}
\begin{equation}\label{eq_BogAns}
\begin{array}{c c}
\chi^{\alpha/\beta}(\rho,\phi,\tau) = \chi^{\alpha/\beta}(\rho,\phi,0) \ \ \ \ \ \ \ \ \ \ \ \ \ \ \ \ \ \ \ \ \ \ \ \ \ \ \ \ \ \ \ \ \ \ \\ \ \ \ \ \ \ \ \ \ \ \ \ + \left[ u^{\alpha/\beta}_{l}(\rho,\phi)e^{iE_{l} \tau} + v^{\alpha/\beta \; *}_{l}(\rho,\phi)e^{iE_{l}^{*} \tau} \right].
\end{array}
\end{equation}
In the above equation, the $u$'s, $v$'s and mode energies $E_{l}$ have been explicitly labeled with their angular momentum quantum number, $l$, to derive the four coupled Bogoliuvbov-de Gennes equations
\begin{equation}\label{eq_bogeqns}
\begin{array}{c c c }
 E_{\kappa,l} u^{\alpha/\beta}_{l} &=& -\left[\partial^2_{\rho} + \frac{1}{\rho}\partial_{\rho} -
  \frac{l^2}{\rho^2}\right]u_{l}^{\alpha/\beta}
                            +2\epsilon|\chi^{\alpha/\beta}|^2 u_{l}^{\alpha/\beta}\\ & & -\epsilon|\chi^{\alpha/\beta}|^2 v_{-l}^{\alpha/\beta} -|\kappa|u_{l}^{\beta/\alpha}\\
 -E_{\kappa,l} v^{\alpha/\beta}_{-l} &=& -\left[\partial^2_{\rho} + \frac{1}{\rho}\partial_{\rho} -
  \frac{l^2}{\rho^2}\right]v_{-l}^{\alpha/\beta}
                            +2\epsilon|\chi^{\alpha/\beta}|^2 v_{-l}^{\alpha/\beta}\\ & & -\epsilon|\chi^{\alpha/\beta}|^2 u_{l}^{\alpha/\beta} -|\kappa|v_{-l}^{\beta/\alpha}
\end{array}
\end{equation}
where $\epsilon = \gamma N_{0}$ and $E_{\kappa,l}$ ($\equiv E_{l}$ in Eq.~\ref{eq_BogAns}) is the energy of the Bogoliubov mode for a particular $\kappa$ and $l$. 

\begin{figure}[tbp]
\includegraphics[width=1.0\columnwidth]{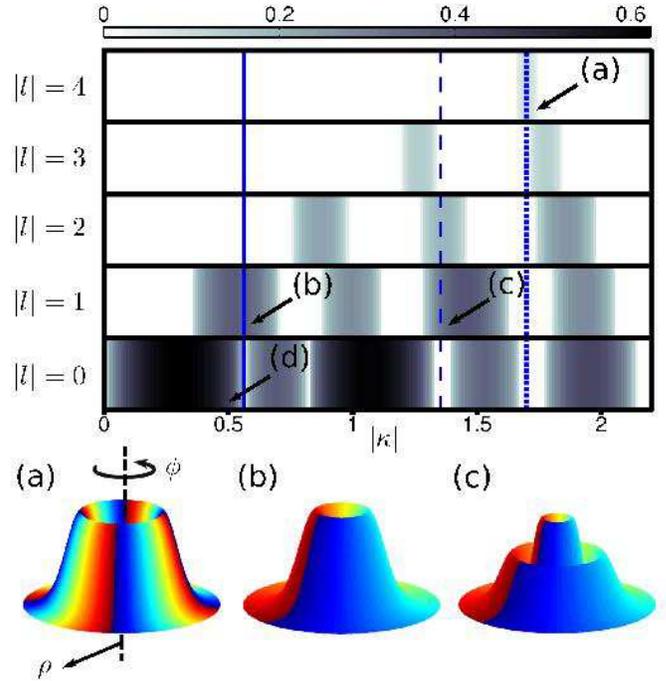} 
\caption{(Color) Upper plot: Imaginary component of ${E_{\kappa,l}}$ (scale shown at top) for angular momentum modes $|l|= 0$ to $4$ as a function of $|\kappa|$. Lower surface plots: three examples of excited modes with (a) $l = 4$ and no radial node ($|\kappa| = 1.7$), (b) $l = 1$ and no radial node ($|\kappa| = 0.56$), (c) $l = 1$ and a single radial node $|\kappa| = 1.35$. The vertical height shows the atomic density of the mode, and the colors denote its phase (blue = 0, green = $\pi$, red = $2\pi$).}
\label{f1}
\end{figure}

The upper panel in Fig. 2 is a gray-scale plot of the imaginary component of $E_{\kappa,l}$ for $|l| = 0$ to $4$ calculated as a function of $|\kappa|$, for a fixed interaction strength $\epsilon=7$. We are interested in complex eigenvalues because they indicate unstable modes which grow exponentially at a rate proportional to the imaginary component. The results are presented in terms of $|l|$ because each Bogoliubov mode has a degenerate pair with equal and opposite angular momentum. Each shaded region in the figure corresponds to a single mode with a particular angular momentum and radial excitation. To illustrate this, we pick three example couplings of $|\kappa|= \left\{1.7,0.56,1.35\right\}$, indicated in the upper panel of Fig. 1 by the vertical $\left\{\right.$dotted, solid, dashed$\left.\right\}$ lines, which pass through shaded regions labeled $\left\{\right.$(a), (b), (c)$\left.\right\}$. The modes corresponding to these shaded regions are shown, also labeled $\left\{\right.$(a), (b), (c)$\left.\right\}$, in the lower half of Fig. 2 as surface plots, whose vertical height denotes the atomic density of the mode, and the colors denote its phase. Mode (a) in the lower panel of Fig. 2 has four quanta of angular momentum ($l=4$), but no radial node. Mode (b) is similar to mode (a), but has only one quantum of angular momentum. As we move to larger values of $|\kappa|$ for a given $|l|$, we find regions of instability corresponding to modes with a greater radial excitation. For example, mode (c) has a higher radial excitation than mode (b) because it has an additional radial density node~\cite{foot3}. 

The positions of the shaded regions in the upper panel of Fig. 2 can be understood by considering the limiting case as $\epsilon \to 0$ (no inter-atomic interactions). In this limit, the shaded regions shrink to points where the coupling strength, $\kappa$, matches the single-particle energies, $n+|l|/2$, where $n$ denotes the number of radial nodes in the corresponding wavefunction. Clearly, in this case, excitations can be degenerate. However, inter-atomic interactions destroy this degeneracy. In general, such interactions shift the shaded regions to higher energy. The size of this shift varies, so that there are certain values of $|\kappa|$ (for example, $|\kappa|= 1.7$, shown by the dotted line in the upper panel of Fig. 2) where only one mode is unstable, and hence we can tune the coupling to excite that mode \emph{alone}. There are also values of $\kappa$ (for example, $|\kappa|= 2.15$) where no modes are unstable, meaning that the $\pi$-state is stable. In addition to shifting the positions of the shaded regions to higher $\kappa$, increasing the interactions from zero also broadens the unstable regions and increases the imaginary component of ${E_{\kappa,l}}$ within them. This illustrates that, although the tunnel coupling provides the energy for the growth of unstable modes, the scattering from the $\pi$-state into the excitation is mediated by the inter-atomic interactions. We find that $\epsilon=7$ is a suitable value to provide reasonable growth rates of excited modes, without causing the unstable regions to become so broad that they overlap, so that specific modes can still be accessed individually.\\

\section{Simulations of the dynamics}

To simulate the growth of the unstable modes, we must model inter-atomic scattering events not captured by the Gross-Pitaevskii equation. Hence we add non-zero angular momentum fluctuations of the form $\sum{C^{(\alpha/\beta)}_{l}r^{|l|}e^{-\rho^2/2}e^{i l\phi}}$ (the single-particle excitations) to $\chi^{\alpha/\beta}$~\cite{fluct}, then evolve Eq. [2] in time. Here, $C^{(\alpha/\beta)}_{l}$ is a random complex number such that $\langle |C^{(\alpha/\beta)}_{l}| \rangle = 0$ and $\langle |C^{(\alpha/\beta)}| \rangle^{2} \approx 10^{-8}\mathcal{O}(N_{0})$. These fluctuations are small enough that they do not raise the temperature significantly from zero in our model.

\subsection{Periodic excitation of a rotating mode ($|\kappa| = 1.7$)} 

Figure 3 shows the results of a simulation for $|\kappa| = 1.7$. For this coupling, only a single mode with $|l| = 4$ [shown by surface plot (a) in the lower panel of Fig. 2] is unstable, as indicated by the vertical dotted line in the upper panel of Fig. 2. Initially, both clouds have a smooth density profile [Fig. 3(a)]. However, as we evolve in time, a macroscopic number of atoms are scattered from the $\pi$-state via the inter-atomic interactions into the $l=4$ mode. To preserve angular momentum, an equal number of atoms transfer into the degenerate $l=-4$ mode, producing the star-like interference pattern with eight density antinodes at $\tau \approx 95$ shown in Fig. 3(b). To quantify the growth of non-zero angular momentum modes, we introduce the quantity $P^{\alpha/\beta}_{l} = 1/2\pi|\int\chi^{\alpha/\beta}(e^{-il\phi} + e^{il\phi})\rho d\rho~d\phi|^2$, which is the projection of the upper or lower DBG onto modes with angular momentum $|l|$. We plot $\text{log}_{10}(P^{\alpha}_{4})$ in Fig. 3(d). The graph shows that the population of the $|l|=4$ mode increases linearly on the logarithmic plot, indicating exponential growth, reaching a maximum of approximately $N_{0}/3$ at $\tau\approx95$. At this point, the growth of the $l=4$ mode halts, and atoms begin transferring back into the $\pi$-state. The population of the $\pi$-state does not reach zero because the stability analysis presented in Fig. 2 assumes that all atoms are in the $\pi$-state, and is therefore no longer valid once a macroscopic number of atoms have entered an excited mode. Eventually, the DBG returns to its original configuration at $\tau \approx 160$ [Fig. 3(c)], at which point the process repeats. This periodic transfer can be maintained because, due to its large angular momentum, the excited mode's peak density is far from the origin, meaning that there is little interaction with the remaining atoms in the $\pi$-state. Consequently, the system is able to supresses particle transfer between the two wells over a large time, as shown in Fig. 3(e), which plots the number imbalance $\eta = (N^{\alpha}-N^{\beta})/2N_{0}$ between the two wells during the simulation~\cite{foot1}.

The width of the unstable $l=4$ mode in Fig. 2 is very narrow, indicating that the formation of the star-like interference pattern is a sharply resonant process. Hence, the growth rate varies extremely sensitively with $\kappa$, reaching zero if $\kappa=1.7$ is changed by only $\sim 2\%$. Similarly, the system is also highly sensitive to the relative phases of the DBGs. For example, when the above simulation is run with an initial relative phase of $0.9 \pi$ between the two clouds, the star-like interference pattern is no longer observed. Moreover, an initial relative phase of $0.99 \pi$ causes a 10 $\%$ change in the growth rate of the star-like pattern. Since this predictable and macroscopic pattern is stable, it could be detected experimentally \textit{in situ} using absorption imaging. 

\subsection{Oscillations in relative atom number and vortex production ($|\kappa| = 0.56$)} 
\label{sec:JO}

\begin{figure}[tbp]
\includegraphics[width=1.0\columnwidth]{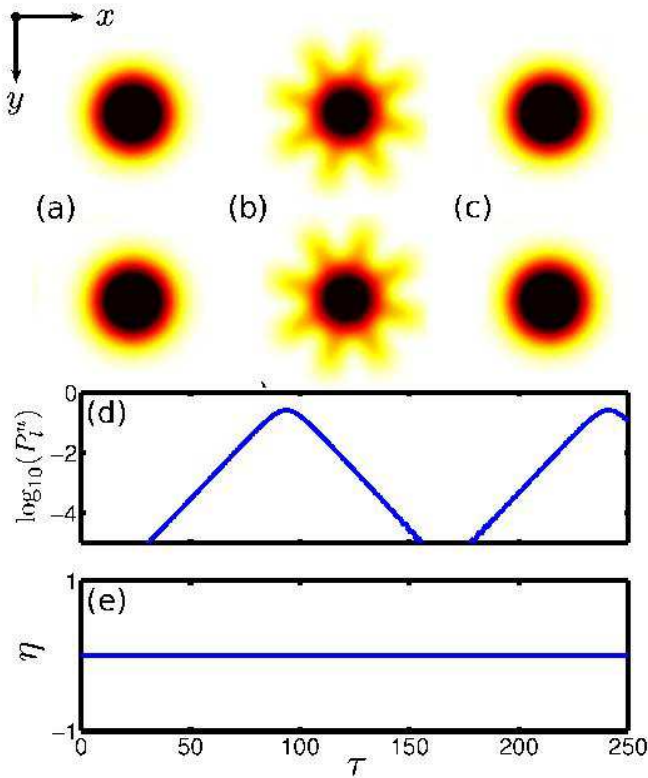} 
\caption{(Color online) Simulation for $|\kappa| = 1.7$. (a)-(c) Atom density plot (black = high) of upper (first row) and lower (second row) DBG at (a) $\tau = 0$, (b) $\tau = 95$ and (c) $\tau = 105$. (d) Logarithmic plot of the projection $P^{\alpha}_{4}$ of the upper DBG onto angular momentum modes with $|l| = 4$. (e) Atom number imbalance, $\eta$, between upper and lower well.}
\label{f2}
\end{figure}

\begin{figure}[tbp]
\includegraphics[width=1.0\columnwidth]{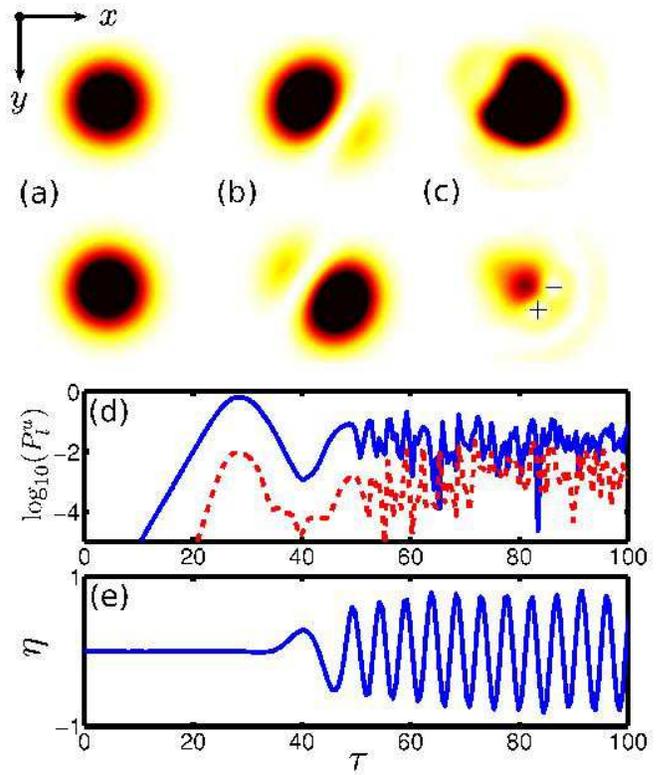} 
\caption{(Color online) Simulation for $|\kappa| = 0.56$. (a)-(c) Atom density plot (black = high) of upper (first row) and lower (second row) DBG at (a) $\tau = 0$, (b) $\tau = 25$ and (c) $\tau = 75$. The positions of vortices are indicated by the black (minus) plus, which denotes (anti) clockwise rotation. (d) Solid (dashed) line: logarithmic plot of the projection $P^{\alpha}_{1}$ ($P^{\alpha}_{2}$) of the upper DBG onto angular momentum modes with $|l| = 1$ ($|l| = 2$). (e) Atom number imbalance, $\eta$, between upper and lower well.}
\label{f3}
\end{figure}

We now tune $|\kappa|$ to $0.56$ to excite the lowest possible rotating mode, as shown by the vertical solid line in the upper panel of Fig. 2. Comparing the surface plots in the lower half of Fig. 2, we see that this mode [Fig. 2(b)] is similar to the one previously excited [Fig. 2(a)], except that it has only one quantum of angular momentum. Consequently, we now expect the formation of an interference pattern with two antinodes. Starting from the smooth atom density profile at $\tau = 0$ [Fig. 4(a)], such an interference pattern begins to form, but the two antinodes have very different peak densities [Fig. 4(b)]. The projection $P^{\alpha}_{1}$ shown by the solid curve in Fig. 4(d) reveals that, as expected from the stability analysis, the population of the $|l|=1$ modes initially grows exponentially. However, as the number of atoms in the $|l| = 1$ modes becomes macroscopic, there is also a significant rise in the projection $P^{\alpha}_{2}$ [dashed curve in Fig. 4(d)], which is not predicted by the stability analysis. We find these unexpected scattering processes when we excite \textit{low} energy modes, which are concentrated close to the center of the trap, and hence collide more frequently with the remaining atoms in the $\pi$-state. Despite these unexpected scattering events, for $10 \lesssim \tau \lesssim 40$ we may still observe one oscillation in the population of the $|l| = 1$ mode in Fig. 4(d). However, at $\tau \approx 50$ the oscillation breaks down, and vortex pairs appear in the density profile, as shown in the lower panel of Fig. 4(c), in which the position and (anti)clockwise rotation of the vortex is indicated by the (minus) plus sign. At the same time, we observe the onset of regular, large amplitude ($\sim 0.8$) oscillations in the number imbalance $\eta$ [plotted in Fig. 4(e)] due to inter-well transfer. These dynamics reflect that, at this low $\kappa$, the system has insufficient energy to populate other rotating modes and, consequently, it redistributes its energy by performing oscillations in relative atom number. This behavior can alternatively be regarded as instability in the symmetric $l=0$ groundstate mode, which occurs within the shaded region labeled (d) in the upper panel of Fig. 2. 

Despite the apparent disruption and the macroscopic population of rotating modes in the simulations presented hitherto, the system has always conserved zero net angular momentum in \textit{each} DBG, as illustrated, for example, by the presence of a vortex anti-vortex pair in Fig. 4(c). However the complete system, defined by Eq. \eqref{eq_gp2d}, only requires that the \textit{total} angular momentum about the $z$ axis be conserved, meaning that each DBG may break its initial chiral symmetry and obtain a net angular momentum, so that $\langle L^{\alpha}_{z} \rangle = -\langle L^{\beta}_{z} \rangle \neq 0$, where $\langle L^{\alpha/\beta}_{z} \rangle$ is the angular momentum of the upper$/$lower DBG. No such mode is predicted fom the stability analysis as the mechanism for spontanous decay of the $\pi$-state involves two atoms with zero initial angular momentum gaining equal and opposite angular momentum via an elastic collision. Since this process is mediated by the non-linear term in Eq. \eqref{eq_gp2d}, it may only occur between atoms in the same well. Consequently, we conclude that the breaking of chiral symmetry is a secondary process, caused by tunneling of angular momentum between the wells, once rotating modes have become populated. We now present an example of this behavior.

\subsection{Counter-rotation of the two clouds ($|\kappa| = 1.35$)} 

\begin{figure}[tbp]
\includegraphics[width=1.0\columnwidth]{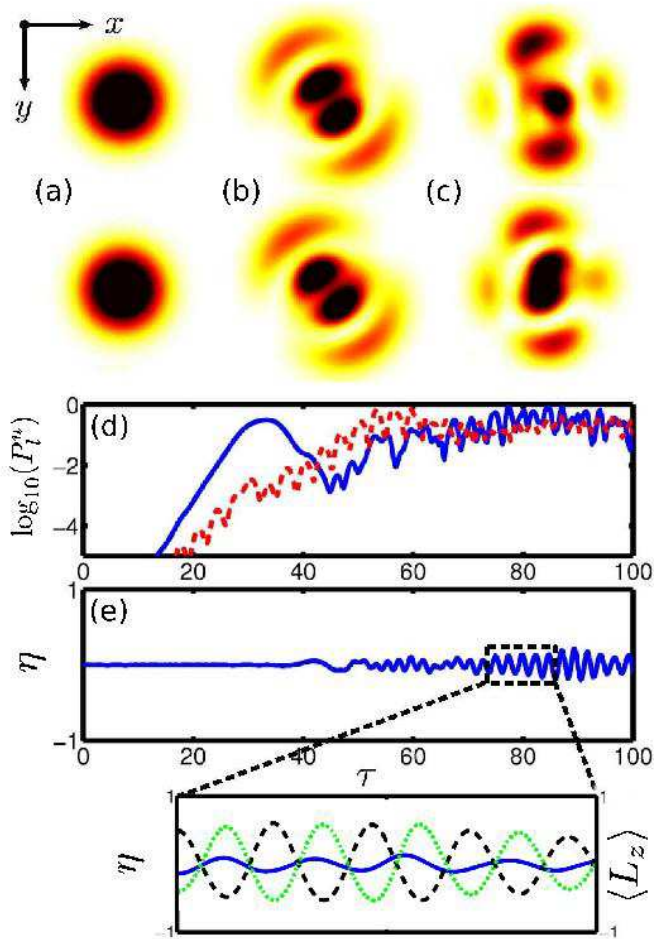} 
\caption{(Color online) Simulation for $|\kappa| = 1.35$. (a)-(c) Atom density plot (black = high) of upper (first row) and lower (second row) DBG at (a) $\tau = 0$, (b) $\tau = 40 $ and (c) $\tau = 100$. (d) Solid (dashed) line: logarithmic plot of the projection $P^{\alpha}_{1}$ ($P^{\alpha}_{2}$) of the upper DBG onto angular momentum modes with $|l| = 1$ ($|l| = 2$). (e) Atom number imbalance, $\eta$, between upper and lower well. Lower plot shows enlargement of curve within dashed box in (e) as a (blue) solid curve with scale shown left, plus $L_z^{\alpha/\beta}$ as a (green) dashed$/$(black) dotted curve with scale shown right.}
\end{figure}

We set $\kappa$ to $1.35$ (dashed line in Fig. 2) in order to excite $|l| = 1$ modes with additional radial energy [shown in surface plot (c) in the lower panel of Fig. 2]. Figure 2 reveals that an $|l| = 2$ mode is also unstable for this coupling, but its imaginary eigenvalue is much smaller (lighter gray in Fig. 2) than the corresponding value for the $|l| = 1$ modes. We would therefore expect the $|l| = 1$ modes to grow faster. Our simulation confirms this: the initially smooth density profile [Fig. 5(a)] transforms into an interference pattern [Fig. 5(b)] with two antinodes as $\phi$ is rotated through $2\pi$, and two antinodes in the $\rho$ direction. The population of this mode [shown by the solid curve in Fig. 5(d)] performs one oscillation for $15 \lesssim \tau \lesssim 45$, but by $\tau \approx 50$ the growing population of the $|l|=2$ modes [shown by dashed line in Fig. 5(d)] has become significant. The occupation of multiple rotating modes disrupts the DBGs' internal structure, but eventually four rough density peaks, encircling the central peak, appear in the atomic density profile, revealing macroscopic population of the $|l|=2$ modes [Fig. 5(c)]. As this occurs, we observe the onset of oscillations in relative atom number, indicated by the periodic fluctuations in $\eta$ shown in Fig. 5(e). The frequency of the oscillations is higher than that shown in Fig. 4(e), because $\kappa$ has been increased~\cite{JJTheo}. Unlike the case presented in section~\ref{sec:JO}, this is accompanied by a transfer of angular momentum between the two DBGs, such that the two DBGs counter-rotate. We quantify this effect in the enlargement of the dashed area in Fig. 5(e), which shows not only $\eta$ (solid curve, left axis), but also $\langle L^{\alpha/\beta}_{z} \rangle$ (dashed$/$dotted curve, right axis). The curves illustrate that, although $\langle L^{\alpha}_{z} \rangle = -\langle L^{\beta}_{z} \rangle$, maintaining the net zero angular momentum of the \emph{whole} system, $\langle L^{\alpha}_{z} \rangle$ and $\langle L^{\beta}_{z} \rangle$ perform periodic oscillations, with a maximum angular momentum difference of $1.6\hbar$. Numerically, we find that the lowest energy counter-rotating mode with $\langle L^{\beta}_{z} \rangle = -\langle L^{\alpha}_{z} \rangle = \hbar$ has an energy of $2N_0(\mu_{0} + 1.3)$. This explains why macroscopic occupation of a counter-rotating mode is not observed in the simulation presented in Fig 3, which has an energy of $2N_0(\mu_{0} + 0.56)$.


\section{Conclusion}

In summary, we have shown that the well known decay mechanism of a three-dimensional $\pi$-state~\cite{RScottInter,RScottInter2,dutton,anderson} is suppressed in tunnel-coupled 2D DBGs. This results in a rich variety of physics, such as spontaneous rotation, vortex formation, and chiral symmetry breaking due to counter-rotation of the two clouds. Our paper builds on previous studies of coupled 1D rings~\cite{annular}, showing that the radial degree of freedom in the 2D disks plays an important role. Firstly, excitations in the radial ($r$) direction appear in the stability diagram, and may be selectively populated and experimentally observed. Secondly, the clouds may develop a complex internal structure, including vortex anti-vortex pairs. Thirdly, the radial position of the peak density, which is fixed in the case of the rings, depends on the angular momentum of the 2D cloud. As a consequence, disruption arises more readily when we populate modes with low angular momentum. 

Since this work is restricted to zero temperature, it remains an open question as to how coupling multiple 2D DBGs will change their dynamics at finite temperature. Finite temperature will introduce a variation of phase across the clouds, which might increase disruption, causing the counter-rotation to occur more readily. It is also possible that coupled systems may provide a link between 2D Kosterlitz-Thouless behavior and three-dimensional Bose-Einstein condensation.



We note that these systems can be realized with current experimental techniques. Similar set ups have already been created by combining a harmonic trap with a one-dimensional optical lattice, to investigate quantum fluctuation-induced localization~\cite{kasevich} and the Kosterlitz-Thouless transition~\cite{hello}. Using this method, values of $\lambda$ up to $370$ have been achieved. A suitable choice of parameters would be, for example, $\omega = 2\pi\times30$Hz, $\lambda = 10$ and $\epsilon = 7$. In this case, $N_{0} \approx 200$ $^{87}$Rb atoms and the conditions for two-dimensionality are satisfied. These parameters are comparable to those in previous experiments~\cite{kasevich}. Using optical lattices to create this system also offers the intriguing possibility of extending our work to spinor DBGs or arrays of coupled DBGs~\cite{kasevich,cornell,thzrad}, which might reveal a crossover from 2D behavior to three-dimensional band dynamics.

Finally, we speculate that this system could potentially be developed as a sensor to detect tiny gravitational or electromagnetic forces, in a similar manner to the production of vortices between merged elongated DBGs~\cite{kettnew,RScottInter,RScottInter2}. In principle, one of the DBGs could be exposed to a tiny force, causing a phase change of $0.01 \pi$, which could then be detected as a change in the growth rate of rotating modes. 

\begin{acknowledgments}
This work is funded by EPSRC-UK. We thank P. Kr{\"u}ger for helpful discussions.
\end{acknowledgments}

\bibliography{biblio}

\begin{thebibliography}{43}
\bibitem[{foo({\natexlab{a}})}]{foot2}
\bibinfo{note}{We assume that there is negligible overlap between the two BECs,
  i.e. $\int^{\infty}_{-\infty} \zeta(z-z_{0}) \zeta(z+z_{0}) dz \approx 0$.}
\bibitem[{foo({\natexlab{b}})}]{foot3}
\bibinfo{note}{There is a small region of instability between regions (b) and
  (c) that has no direct analogy in the non-interacting case.}
\bibitem[{foo({\natexlab{c}})}]{foot1}
\bibinfo{note}{Even though the relative phase between the rotating and
  non-rotating atoms in different clouds may be a non-integer multiple of
  $\pi$, particle transfer cannot occur between these states because their
  angular momenta do not match.}
\end{thebibliography}

\end{document}